\documentclass[%twocolumn,
aps,prd,   
               preprintnumbers,numbers,sort&compress,
               nofootinbib,
                            showpacs,
               colorlinks,
               linkcolor=blue,
               citecolor=blue]{revtex4-1}

   \newcommand{\exclude}[1]{}
\newcommand{\be}{\begin{eqnarray}}
\newcommand{\ee}{\end{eqnarray}}

\usepackage{graphicx,amsmath,amssymb,bm}

\usepackage{psfrag}
 \usepackage{feynmp}
 \usepackage{hyperref}

\usepackage{graphicx}
\usepackage{subcaption}

\newcommand{\beq}{\begin{equation}}
\newcommand{\eeq}{\end{equation}}

\def\ra{\rangle}
\def\la{\langle}

\begin{document}

\title{New  mechanism  producing  axions  in the AQN model
   and how the CAST can discover them} 
\author{H. Fischer}
\email{Horst.Fischer@cern.ch}
 \affiliation{University of Freiburg, Freiburg, 79104 Germany}
 \author{X.Liang}
 \email{xunyul@phas.ubc.ca}
\affiliation{Department of Physics \&  Astronomy, University of British Columbia, V6T1Z1
Vancouver,  Canada}
  \author{Y. Semertzidis}
  \email{yannis@ibs.re.kr}
 \affiliation{Center for Axion and Precision Physics Research, Institute for Basic Science, Daejeon, 34141, Republic of Korea\\
 and Department of Physics, KAIST, Daejeon 34141, Republic of Korea}
 \author{A.  Zhitnitsky}
 \email{arz@physics.ubc.ca}
  \affiliation{Department of Physics \&  Astronomy, University of British Columbia, V6T1Z1
Vancouver,  Canada}
  \author{ K. Zioutas}
  \email{Konstantin.Zioutas@cern.ch}
    \affiliation{Physics Department, University of Patras, Patras, GR 26504 Greece}

  \begin{abstract} 
 We advocate the idea that   there  is a fundamentally new mechanism  for the axion production in the Sun, which has never been discussed previously in the literature. This  novel mechanism of the axion production is based on the so-called Axion Quark Nugget (AQN) 
Dark Matter Model.    These 
  axions will be produced in addition to well studied  axions emitted    due to the Primakoff effect. 
    The AQN     model was originally invented as a natural explanation of  the 
observed ratio $\Omega_{\rm dark} \sim   \Omega_{\rm visible}$ when the DM and visible matter densities assume the same order of magnitude  values, irrespectively to the axion mass $m_a$ or initial misalignment angle $\theta_0$.
 This model, without adjustment of any parameters,  reproduces reasonably the intensity of the extreme UV  (EUV) radiation from the solar corona 
  as a result of the AQN annihilation events with the solar material.    This extra energy released in corona  represents a resolution, within AQN framework, a long standing puzzle known in the literature as the ``solar corona heating mystery".  The same annihilation events also produce the relativistic  axions. This represents a new mechanism of the axion production, and constitutes the main subject of this work. 
  The flux of these axions is unambiguously fixed  in this model and expressed in terms of the  EUV luminosity from corona. We also compute the spectral properties  of these  axions and make few comments on the potentials for the   discovery of these solar axions by the  upgraded CAST (CERN Axion Solar Axion)  experiment. 
 \end{abstract}
 \maketitle
%\flushbottom
\section{Introduction}
The Peccei-Quinn (PQ) mechanism accompanied  by the axions remains the most compelling explanation of the strong CP problem, see original papers 
  \cite{axion,KSVZ,DFSZ} and  
 recent reviews  \cite{vanBibber:2006rb, Asztalos:2006kz,Sikivie:2008,Raffelt:2006cw,Sikivie:2009fv,Rosenberg:2015kxa,Marsh:2015xka,Graham:2015ouw,Ringwald:2016yge} on the subject. We refer to the review  articles for the discussions and analysis  on the  recent activities in the field of the axion searches   by a numerous number of different groups using very   different instruments.
 
 For the purposes of the present work it is sufficient  to mention  that the   conventional dark matter galactic  axions are  produced 
 due to the misalignment mechanism \cite{misalignment} when the cosmological field $\theta(t)$ oscillates and emits cold axions before it settles down at 
 its final destination $\theta_{\rm final}=0$.
 Another mechanism is due  to the   decay of the topological objects \cite{Chang:1998tb,Hiramatsu:2012gg,Kawasaki:2014sqa,Fleury:2015aca,Klaer:2017ond}.   There is a number of uncertainties and  remaining discrepancies in the corresponding estimates. We shall not comment on these subtleties\footnote{\label{DM}
  According to the most recent computations presented in ref.\cite{Klaer:2017ond}, the axion contribution to $\Omega_{\rm DM}$ as a result of decay of the  topological objects can saturate the observed DM density today if the axion mass is  in the range $m_a=(2.62\pm0.34)10^{-5} {\rm eV}$, while the earlier estimates suggest that the saturation occurs at a larger axion mass. One should also emphasize that the computations  \cite{Chang:1998tb,Hiramatsu:2012gg,Kawasaki:2014sqa,Fleury:2015aca,Klaer:2017ond} have been performed with assumption that PQ symmetry was broken after inflation.}
 by referring to the original   papers \cite{Chang:1998tb,Hiramatsu:2012gg,Kawasaki:2014sqa,Fleury:2015aca,Klaer:2017ond}. It is important that  in both cases
 the produced  axions are non-relativistic particles with typical $v_{\rm axion}/c\sim 10^{-3}$, and their contribution to the dark matter density scales as $\Omega_{\rm axion}\sim m_a^{-7/6}$. This scaling  unambiguously implies that the axion mass must be fine-tuned  $m_a\simeq 10^{-5}$ eV
 to  saturate the DM density today, see footnote {\ref{DM},  while larger axion mass will contribute very little to $\Omega_{\rm DM}$.
  The cavity type experiments have a potential to discover these  non-relativistic axions. 
 
 Axions can be also produced as a result of the Primakoff effect in a stellar plasma at high temperature \cite{Sikivie:1983ip}. These axions are ultra-relativistic as the typical average energy of the axions emitted by the Sun is $\la E\ra =4.2$ keV, see \cite{Andriamonje:2007ew}. 
Today the most sensitive broadband searches for solar axions come from the helioscope CAST at CERN   \cite{Andriamonje:2007ew}.

 \exclude{  the International Axion Observatory (IAXO) \cite{Irastorza:2011gs,Armengaud:2014gea},  the Troitsk Axion Solar Telescope Experiment (TASTE) \cite{TASTE}, or similar instruments.}

The main goal of the present work is to argue that there is a fundamentally novel mechanism of the  axion production  in the Sun. 
This mechanism is rooted to  the so-called axion quark nugget (AQN) dark matter model. We overview the basic ideas of this model in next section \ref{sec:QNDM}. Meantime we want to make two important comments related to this model in the context of the axion physics. 

First basic consequence of this model is as follows. 
We already mentioned about  two mechanisms on    energy transfer from the original cosmological axion field $\theta(t)$    to the DM non-relativistic axions as a result of the misalignment mechanism 
  and  decay of the topological defects. The key element of the present work is that in addition to these two well established processes    there is one more path  how the original cosmological  field $\theta(t)$  can  transfer its energy to    the axions.
  This mechanism is based on the idea that the AQNs might be  formed during the same QCD epoch when two other processes of the axion production were operational. This process of formation  inevitably  includes    the closed    $N=1$   axion domain walls as an essential part of the construction. 
      The corresponding axion energy (hidden in the form of the   axion domain wall) is not available unless  the AQN itself gets annihilated and destroyed, in which case the  axion energy  will be released into the space in the form of the  free propagating axions which can be observed on Earth. 
  
  The second important comment     is that these emitted axions   will be released with relativistic (but not ultra-relativistic) velocities  with 
typical values $v_{\rm axion}^{\rm AQN}\simeq 0.5 c$. These features should be   contrasted with conventional galactic non-relativistic axions    $v_{\rm axion}\sim 10^{-3}c$ and solar ultra-relativistic axions with typical energies 
$\la E\ra =4.2$ keV. 

 We highlight  the basic logic and  the ideas    of the AQN dark matter  in next section \ref{sec:QNDM}.
 Now we want to present few observational hints in context of the solar physics which apparently  support this generic AQN proposal. 
 
  The AQN model   was invented long ago  \cite{Zhitnitsky:2002qa} (though a specific formation mechanism   of the nuggets was 
 developed in  much more recent papers \cite{Liang:2016tqc,Ge:2017ttc,Ge:2017idw})
  as a natural explanation of  the observed ratio $ \Omega_{\rm dark}\sim \Omega_{\rm visible}$.  The similarity between  dark matter $ \Omega_{\rm dark}$ and the visible matter $\Omega_{\rm visible}$  densities   strongly suggests that both types of matter  have been formed  during the same cosmological epoch, which must be the QCD transition as the baryon mass $m_p$ which represents the visible portion of the matter $\Omega_{\rm visible}$ is obviously  proportional to $ \Lambda_{\rm QCD}$, while the contribution related to the E\&W    physics proportional to the quark mass $\sim m_q$ represents only a
minor contribution to the proton mass.  

In context of the present work  the argument supporting the AQN model goes as follows.  It has been known  for quite some time   that the  total intensity of the  observed EUV (Extreme Ultra Violet) and (soft) x-ray radiation (averaged over time)   can be estimated as follows,
\be
\label{estimate}
   L_{\odot ~  (\rm from ~Corona)}  \sim 10^{30}\cdot \frac{\rm GeV}{\rm  s} \sim 10^{27}  \cdot  \frac{\rm erg}{\rm  s},
 \ee
 which represents (since 1939) the renowned  ``the solar corona heating puzzle", see e.g. a general review \cite{Klimchuk:2005nx}
 on the subject  and also  Ref. \cite{LZ:2003} with analysis of some specific features related to present work.
  The observation (\ref{estimate}) implies that the corona
 has the temperature  $T\simeq 10^6$K which is 100 times hotter than the surface temperature of the Sun, and conventional astrophysical sources fail to explain the EUV and soft x ray radiation from corona \cite{Klimchuk:2005nx,LZ:2003}.
    
    It turns out that if one estimates   the extra energy being produced within the AQN dark matter scenario     one obtains
 the total extra energy $\sim 10^{27}{\rm erg}/{\rm  s}$    which 
surprisingly reproduces  (\ref{estimate})   for  the   observed EUV and soft x-ray intensities  \cite{Zhitnitsky:2017rop}. One should add that the estimate   $\sim 10^{27}{\rm erg}/{\rm  s}$  for extra energy  is derived  exclusively in terms of known  dark matter density $\rho_{\rm DM} \sim 0.3~ {\rm GeV cm^{-3}}$ and dark matter  velocity $v_{\rm DM}\sim 10^{-3}c $ surrounding the sun  
 without adjusting any  parameters of the model, see  section \ref{AQN-flares}  with relevant estimates.  The recent numerical Monte Carlo simulations carried out in \cite{{Raza:2018gpb}} strongly support this estimation. 
 We  interpret this ``numerical coincidence"  as an additional indication supporting the AQN model.  Our original remark relevant for the present work  is that if one accepts the explanation   \cite{Zhitnitsky:2017rop,Raza:2018gpb}  that the solar corona heating puzzle is resolved within AQN scenario then the axion flux will be unambiguously  fixed in terms of the EUV observed luminosity (\ref{estimate}) as 
 the axion field  represents the crucial element in the AQN construction. 
 
 Another inspiring observation supporting the AQN scenario in  the context of the present studies can be explained as follows.
 It was recently claimed in ref. \cite{Zioutas}   that a number of highly unusual phenomena   observed in the solar atmosphere can be explained by  the gravitational lensing of ``invisible" streaming matter towards the Sun, see also \cite{LZ:2003}. The phenomena include,  but not limited to  such irradiation as the EUV emission, frequency of X and M flare occurrences, etc. 
Naively, one should not  expect    any correlations between the   flare occurrences, the intensity of the EUV radiation,    and the position of the planets. 
Nevertheless,     the analysis  \cite{Zioutas} obviously demonstrates that this naive expectation is not quite correct. At the same time,
the emergence of  such correlations  within AQN framework is a quite  natural effect. This is because the dark matter AQNs can play the role of the ``invisible" matter in ref. \cite{Zioutas}, which triggers otherwise unexpected solar activity sparking also the large flares \cite{Zhitnitsky:2018mav}. Therefore, the observation of the correlation between the EUV intensity and frequency of the flares  can be considered as an additional supporting argument  of  the AQN related dark matter explanation of the observed EUV irradiation  (\ref{estimate}), 
because both effects are originated from the same dark matter AQNs. As a direct consequence of this relation we expect that the intensity of 
  the axion emission from the Sun (which always accompanies the EUV emission) will be also correlated  with the position of the planets.

 The paper is organized as follows. In next section \ref{sec:QNDM} we overview the AQN model by paying special attention to the astrophysical and cosmological consequences of this specific dark matter  model. In section \ref{AQN-flares} we highlight  the basic arguments of ref.  \cite{Zhitnitsky:2017rop} advocating the idea that the annihilation events of the antinuggets with the solar material can be interpreted as the   nanoflares conjectured  by Parker      long ago.   Precisely these annihilation events emit the axions and we compute the intensity and spectral properties of these axions in Section \ref{spectrum}. Finally, in Section \ref{design} we highlight the basic ideas of the design of a new detector and comment on possible potential of the discovery of these axions emitted from the solar corona.    We conclude in Section \ref{conclusion} with few thoughts on the future development of  the solar axion searches.

\section{Axion Quark Nugget (AQN) dark matter model}\label{sec:QNDM}
The axion field plays a key role in the construction. Therefore, we would like to make a short overview 
of this model with emphasis  on the role of the axion field 
and related astrophysical consequences of this proposal. 

The idea that the dark matter may take the form of composite objects of 
standard model quarks in a novel phase goes back to quark nuggets  \cite{Witten:1984rs}, strangelets \cite{Farhi:1984qu}, nuclearities \cite{DeRujula:1984axn},  see also review \cite{Madsen:1998uh} with large number of references on the original results. 
 The AQN model in the title of this section stands for the axion quark nugget   model \cite{Zhitnitsky:2002qa}   to emphasize on essential role of the axion field in the construction and to avoid confusion with earlier models    \cite{Witten:1984rs,Farhi:1984qu,DeRujula:1984axn,Madsen:1998uh} mentioned above.
The  AQN  model   is drastically different from previous similar proposals in two key aspects:\\
1. There is an  additional stabilization factor in the AQN  model provided    by the $N=1$ {\it  axion domain walls}
  which are copiously produced during the QCD transition.
   \\  
  2. The AQN  could be 
made of matter as well as {\it antimatter} in this framework as a result of separation of charges, see  recent papers \cite{Liang:2016tqc, Ge:2017ttc,Ge:2017idw} with large number of technical details.  

To recapitulate these two important ingredients: the axions  play a dual  role in construction of the AQNs as they provide an additional pressure to stabilize the nuggets and also play the role of the source (through the axion field) which breaks $\cal{P}$ and $\cal{CP}$ symmetries during the QCD transition.
Precisely these $\cal{P}$ and $\cal{CP}$  violating processes are responsible for the separation of charges, leading to the fundamental and very generic consequence of this framework expressed as $\Omega_{\rm dark}\sim \Omega_{\rm visible}$.
The key role of the axion field  at present  epoch  manifests  itself as a substantial  contribution ($\sim 1/3$) to  the total nugget's  mass \cite{Ge:2017idw}.

The basic idea of  the AQN  proposal can be explained   as follows: 
It is commonly  assumed that the Universe 
began in a symmetric state with zero global baryonic charge 
and later (through some baryon number violating process, the so-called baryogenesis) 
evolved into a state with a net positive baryon number. As an 
alternative to this scenario we advocate a model in which 
``baryogenesis'' is actually a charge separation process 
when  the global baryon number of the Universe remains 
zero. In this model the unobserved antibaryons come to comprise 
the dark matter in the form of dense nuggets of quarks and antiquarks in colour superconducting (CS) phase.  
  The formation of the  nuggets made of 
matter and antimatter occurs through the dynamics of shrinking axion domain walls, see
original papers \cite{Liang:2016tqc,Ge:2017ttc,Ge:2017idw} with many technical  details. 

 The  nuggets, after they formed,  can be viewed as the  strongly interacting and macroscopically large objects with a  typical  nuclear density 
and with a typical size $R\sim (10^{-5}-10^{-4})$cm determined by the axion mass $m_a$ as these two parameters are linked, $R\sim m_a^{-1}$.
This  relation between the size of nugget $R$ and the axion mass $m_a$  is a result of the equilibration between the axion domain wall pressure and the Fermi pressure 
of  the dense quark matter  in CS phase. 
One can easily    estimate a typical  baryon charge $B$ of  such  macroscopically large objects as the typical density of matter in   CS phase  
is only few times the   nuclear density. 
However, it is important to emphasize that there are strong constraints on the    allowed window for the axion mass,  which can be represented as follows $10^{-6} {\rm eV}\leq m_a \leq 10^{-2} {\rm eV}$.
 This axion window corresponds to the range of the nugget's baryon charge $B$ which   largely overlaps  with all presently available and independent constraints on such kind of dark matter masses and baryon charges 
 \beq
 \label{B-range}
 10^{23}\leq |B|\leq 10^{28}, 
 \eeq
 see e.g. \cite{Jacobs:2014yca,Lawson:2013bya} for review.   The corresponding mass ${\cal M}$ of the nuggets  can be estimated as ${\cal M}\sim m_pB$, where $m_p$ is the proton mass.  

 This  model is perfectly consistent with all known astrophysical, cosmological, satellite and ground based constraints within the parametrical range for 
 the mass ${\cal M}$ and the baryon charge $B$ mentioned   above (\ref{B-range}). It is also consistent with known constraints from the axion search experiments. Furthermore, there is a number of frequency bands where hints for excess of emission exist, but could not be explained by conventional astrophysical sources. Our comment here is that this model may explain some portion, or even entire excess of the observed radiation in these frequency bands, see short review \cite{Lawson:2013bya} and additional references at the end of this section.

Another key element of this model is   the coherent axion field $\theta$ which is assumed to be non-zero during the QCD transition in early Universe.
       As a result of these $\cal CP$ violating processes the number of nuggets and anti-nuggets 
      being formed would be different. This difference is always of order  one effect   \cite{Liang:2016tqc,Ge:2017ttc,Ge:2017idw} irrespectively to the parameters of the theory, the axion mass $m_a$ or the initial misalignment angle $\theta_0$. As a result of this disparity between nuggets and anti nuggets   a similar disparity would also emerge between visible quarks and antiquarks.  
       This  is precisely  the reason why the resulting visible and dark matter 
densities must be the same order of magnitude \cite{Liang:2016tqc,Ge:2017ttc,Ge:2017idw}
\be
\label{Omega}
 \Omega_{\rm dark}\approx \Omega_{\rm visible}
\ee
as they are both proportional to the same fundamental $\Lambda_{\rm QCD} $ scale,  
and they both are originated at the same  QCD epoch.
  If these processes 
are not fundamentally related the two components 
$\Omega_{\rm dark}$ and $\Omega_{\rm visible}$  could easily 
exist at vastly different scales. 
 
 Another fundamental ratio (along with $\Omega_{\rm dark} \approx  \Omega_{\rm visible}$   discussed above)
is the baryon to photon ratio at present time
\be
\label{eta}
\eta\equiv\frac{n_B-n_{\bar{B}}}{n_{\gamma}}\simeq \frac{n_B}{n_{\gamma}}\sim 10^{-10}.
\ee
If the nuggets were not present after the phase transition the conventional baryons 
and antibaryons would continue to annihilate each other until the temperature 
reaches $T\simeq 22$ MeV when density would be 9 orders of magnitude smaller 
than observed (\ref{eta}). This annihilation catastrophe, normally thought   to be  resolved   as a result of  ``baryogenesis" as formulated by Sakharov \cite{Sakharov}.  

In our proposal (in contrast with conventional frameworks on baryogenesis) the annihilation stops because the extra anti-baryon charge is hidden in the antinuggets, while the total baryon charge of the Universe remains zero at all times. 
The ratio (\ref{eta}) in the AQN framework is determined by a single parameter with a typical  QCD scale, the formation temperature $T_{\rm form}\approx 40~ {\rm MeV}$, slightly lower than the critical temperature $T_c$ of the gap in CS phase. This temperature is defined by a moment in  evolution of the Universe  when the nuggets and antinuggets basically have completed  their formation and not much annihilation would occur at lower  temperatures $T \leq T_{\rm form}$, see     
 \cite{Liang:2016tqc,Ge:2017ttc} for the details. 

  Unlike conventional dark matter candidates, such as WIMPs 
(Weakly interacting Massive Particles) the dark-matter/antimatter
nuggets are strongly interacting and macroscopically large objects,  as we already mentioned. 
However, they do not contradict any of the many known observational
constraints on dark matter or
antimatter    in the Universe due to the following  main reasons~\cite{Zhitnitsky:2006vt}:
  They carry  very large baryon charge 
$|B|  \gtrsim 10^{23}$, and so their number density is very small $\sim B^{-1}$.  
 As a result of this unique feature, their interaction  with visible matter is rare, and 
therefore, the nuggets  perfectly qualify  as  DM  candidates. Furthermore, 
  the quark nuggets have  very  large binding energy due to the   large    gap $\Delta \sim 100$ MeV in  CS phases.  
Therefore, the baryon charge is so strongly bounded in the core of the nugget that  it  is not available to participate in big bang nucleosynthesis
(\textsc{bbn})  at $T \approx 1$~MeV, long after the nuggets had been formed.

  It should be noted that the galactic spectrum 
contains several excesses of diffuse emission the origin of which is unknown, the best 
known example being the strong galactic 511~keV line. If the nuggets have the  average  baryon 
number in the $\langle B\rangle \sim 10^{25}$ range they could offer a 
potential explanation for several of 
these diffuse components.  
\exclude{(including 511 keV line and accompanied   continuum of $\gamma$ rays in 100 keV and few  MeV ranges, 
as well as x-rays,  and radio frequency bands). }
It is important to emphasize that a comparison between   emissions with drastically different frequencies in such  computations 
 is possible because the rate of annihilation events (between visible matter and antimatter DM nuggets) is proportional to 
one and the same product    of the local visible and DM distributions at the annihilation site. 
The observed fluxes for different emissions thus depend through one and the same line-of-sight integral 
\be
\label{flux1}
\Phi \sim R^2\int d\Omega dl [n_{\rm visible}(l)\cdot n_{DM}(l)],
\ee
where $R\sim B^{1/3}$ is a typical size of the nugget which determines the effective cross section of interaction between DM and visible matter. As $n_{DM}\sim B^{-1}$ the effective interaction is strongly suppressed $\sim B^{-1/3}$. The parameter $\la B\ra\sim 10^{25}$  was fixed in this  proposal by assuming that this mechanism  saturates the observed  511 keV line   \cite{Oaknin:2004mn, Zhitnitsky:2006tu}, which resulted from annihilation of the electrons from visible matter and positrons from antinuggets.   Other emissions from different frequency bands  are expressed in terms of the same integral (\ref{flux1}), and therefore, the  relative  intensities  are unambiguously and completely determined by internal structure of the nuggets which is described by conventional nuclear physics and basic QED, see 
short overview  \cite{Lawson:2013bya} with references on specific computations of diffuse galactic radiation  in different frequency bands. 

Finally we want to mention that the recent EDGES (Experiment to Detect the Global Epoch of reionization Signatures) observation of a stronger than 
anticipated 21 cm absorption  \cite{Bowman:2018yin} can find an explanation within the AQN framework as recently advocated in 
\cite{Lawson:2018qkc}.  The basic idea is that the extra   thermal 
emission from AQN  dark matter at early times  produces the required intensity (without  adjusting  any parameters) to explain the recent EDGES observation.

  \section{ AQNs as the corona's heaters}\label{AQN-flares} 
    Our goal here is to    overview the basic parameters related to the AQNs entering the solar atmosphere from outer space. 
  The impact parameter for capture and crash of the nuggets by the Sun can be estimated as
  \be
  \label{capture}
  b_{\rm cap}\simeq R_{\odot}\sqrt{1+\gamma_{\odot}}, ~~~~ \gamma_{\odot}\equiv \frac{2GM_{\odot}}{R_{\odot}v^2},
  \ee
  where $v\simeq 10^{-3}c$ is a typical velocity of the nuggets.    
    Assuming that $\rho_{\rm DM} \sim 0.3~ {\rm GeV cm^{-3}}$ and using the capture impact parameter (\ref{capture}), one can estimate 
  the total energy flux due to the complete annihilation of the nuggets,
   
  \be
  \label{total_power}
   L_{\odot ~  \rm (AQN)}\sim 4\pi b^2_{\rm cap}\cdot v\cdot \rho_{\rm DM}   
  \simeq 3\cdot 10^{30} \cdot \frac{\rm GeV}{\rm  s}\simeq 4.8 \cdot 10^{27} \cdot  \frac{\rm erg}{\rm  s}, 
  \ee
   where we substitute  constant $v\simeq 10^{-3}c$  to simplify numerical  analysis.  
   This is obviously an order of magnitude estimate as we ignore a large number of factors of order one\footnote{\label{neutrinos}For example, we ignore that only antinuggets, not nuggets generate the energy (\ref{total_power}). Furthermore, a large amount of annihilation energy will be emitted in form of the neutrinos. In addition, even $E\&M$ energy released as a result of annihilation might be radiated in different energy spectrum,  not necessary 
   in form of EUV radiation. Finally, approximately $1/3$ of the energy will be emitted in form of  the axions as we shall argue in next section \ref{spectrum}.}.
    Nevertheless,  this order of magnitude estimate is very suggestive as it roughly coincides with the observed total EUV  energy output   from corona (\ref{estimate}) 
   representing $\sim (10^{-7}-10^{-6})$ portion of the total solar luminosity. 
    Precisely this ``accidental  numerical coincidence" was the main motivation   to put forward the idea \cite{Zhitnitsky:2017rop}
 that  (\ref{total_power}) represents a new source of energy feeding the EUV and soft x-ray radiation. The numerical simulations \cite{Raza:2018gpb} strongly support the estimate (\ref{total_power}) and entire picture of the framework. In particular, the numerical studies \cite{Raza:2018gpb} show that the annihilation events mostly occur at the altitude close to 2000 km where the temperature of the plasma $T\simeq 10^6 {\rm K}$. Therefore, it is  quite natural to expect that the most photons emitted from the annihilation events in the environment will have the energies in EUV and soft x-rays bands. 
       
     The basic claim  of  \cite{Zhitnitsky:2017rop,Raza:2018gpb} is that  the annihilation events of the antinuggets, which  generate  huge amount of  energy (\ref{total_power}) can be  identified with the ``nanoflares"  conjectured by Parker long ago  \cite{Parker}. 
    In most studies the term ``nanoflare" describes a generic burst-like event for any impulsive energy release on a small scale, without specifying its cause. In other words, in most studies the hydrodynamic consequences of impulsive heating (due to the nanoflares) have been used without discussing their nature, see review papers \cite{Klimchuk:2005nx,Klimchuk:2017}.  The novel element of  ref.  \cite{Zhitnitsky:2017rop} is that the nature of the nanoflares was specified as 
   annihilation events  of the dark matter 
   particles within AQN framework, i.e.
              \be
  \label{identification}
  {\rm nanoflares}\equiv {\rm AQN~ annihilation~ events},
  \ee  
  in which case the observed intensity of the EUV (\ref{total_power}) is determined by the DM density $ \rho_{\rm DM} $ in the solar system. 
      The main arguments  of  \cite{Zhitnitsky:2017rop,Zhitnitsky:2018mav,Raza:2018gpb}  supporting the identification (\ref{identification}) 
      and the basic picture in general are:
       
  1. In order   to reproduce the measured  radiation loss, the observed range of nanoflares  needs to be extrapolated   from sub-resolution events with energy $3.7\cdot 10^{20}~{\rm erg}$ to the observed events  interpolating between   $(3.1\cdot 10^{24}  - 1.3\cdot 10^{26})~{\rm erg}$.    This energy window corresponds to the 
 (anti)baryon charge of the nugget $ 10^{23} \leq |B|\leq  4\cdot 10^{28}$  which largely  overlaps with allowed window   (\ref{B-range}) for AQNs reviewed  in section \ref{sec:QNDM}. This  is a highly nontrivial consistency check for the proposal 
 (\ref{identification}) as the window (\ref{B-range}) comes from a number of different and independent  constraints extracted from 
 astrophysical, cosmological, satellite and ground based observations.  The window  (\ref{B-range}) is also consistent with known constraints from the axion search experiments within the AQN framework. Therefore, the overlap between these  two fundamentally different entities represents  a highly nontrivial consistency check of the proposal  (\ref{identification}).
 
2. The corresponding $E\&M$ radiation  is expected to be mostly  in form of the EUV and soft x-ray emissions because the annihilation events of the AQNs mostly occur at the altitude  around 2000 km with a typical width around few hundred kilometres and the typical temperature $T\simeq 10^6$ K. This  extra energy injection (\ref{total_power}) represents in our framework the explanation of the unusual features of the so-called transition region (TR)  when   the temperature of the plasma  experiences  some  drastic  changes  by two orders of magnitude on the length scales of few hundred kilometres. 

3.  Our next argument  goes as follows.     
The nanoflares are distributed very  uniformly in quiet  regions, in contrast with micro-flares and flares 
 which are much more energetic and occur exclusively in active areas.  It  is consistent with our  identification (\ref{identification})  as the  anti-nugget annihilation events   should be present in all areas irrespectively to the activity of the Sun.  At the same time the   flares, triggered by the AQNs as suggested in \cite{Zhitnitsky:2018mav},  are originated in the  active zones, and therefore cannot be uniformly distributed.

 4. The observed  Doppler shifts (corresponding to velocities $ 250-310$ km/s)   and the observed line width in OV  of $\pm 140$ km/s far exceed the thermal ion velocity which is around 11 km/s as discussed in  \cite{Zhitnitsky:2017rop}.
 These observed  features    can be  easily understood  within the AQN scenario.  Indeed, the typical velocities of the nuggets entering the solar corona is about $ \sim  618 ~{\rm km/s}$, the escape velocity of the sun. Therefore, it is perfectly consistent with observations of the  very large Doppler shifts and related broadenings of  the line widths.  Typical time scales of  the nanoflare events, of order of $(10^1-10^2)$ sec are also consistent with estimates  \cite{Zhitnitsky:2017rop}.
 
 5. It has been observed  \cite{x-ray} that ``the pre-flare enhancement propagates from the higher levels of the corona into the lower corona and chromosphere."  
It is perfectly consistent with our proposal as the dark matter AQNs   enter the solar atmosphere from outer space. Therefore, they first enter the higher levels of the corona where they generate the shock wave, before they reach   chromosphere in $\tau\sim (10-10^2)~ {\rm sec}$. 

6.  It has been claimed in   \cite{Shibata:2007,Shibata:2016} that the observations show the ``ubiquitous presence of chromospheric anemone jets outside of sunspots...". In our framework the jet-like structure is a direct consequence of the AQNs entering the solar atmosphere when the nuggets generate the shock waves  with large Mach number $M\sim 10$ which represents a typical jet-like structure  \cite{Zhitnitsky:2018mav}.
One should emphasize here that the most of these events are sub-resolution events which are well below the instrumental threshold    $\sim 3\cdot 10^{24}~ {\rm erg}$, see item 1 above.

\section{ Intensity and the Spectrum of the Solar Axions\label{spectrum}}
In Section \ref{sec:QNDM} we explained  that the axion field is  the key element in the  AQN construction.  
In  Section \ref{AQN-flares} we argued that the AQNs may serve as the heaters of the  corona. In this Section we estimate the intensity and spectral properties of the axions which  will be  inevitably produced as a result of the 
annihilation and complete disintegration of the antinuggets in the solar corona.  
\subsection{Intensity}
  The axions  play a key role in construction of the AQNs as they provide an additional pressure to stabilize the nuggets, see section  \ref{sec:QNDM}
  for review. 
   The corresponding axion contribution into the total nugget's energy density has been computed in \cite{Ge:2017idw}, see the red curve in Fig. 9 in  \cite{Ge:2017idw}. Depending on parameters the axion's contribution to the nugget's mass  represents about 1/3 of the total mass.
   It implies that this entire energy will be radiated in form of the free propagating axions. 
 This energy  can be expressed  in terms of the axion luminosity from the Sun as   follows 
  \be
  \label{axions_rate}
  L_{\odot ~  \rm (axion)}\sim  \frac{1}{3} L_{\odot ~  \rm (AQN)}\simeq 1.6 \cdot 10^{27} \cdot  \frac{\rm erg}{\rm  s}
  \ee
  where $L_{\odot ~  \rm (AQN)}$ is given by (\ref{total_power}).
   The corresponding  axion flux measured on Earth 
     can be computed  as follows
  \be
  \label{axions}
  \Phi_{\rm axions}\sim \frac{L_{\odot ~  \rm (axion)}}{4\pi \la E_a\ra D^2_{\odot}}\sim 3\cdot 10^{16}\frac{1}{\rm cm^2 ~s}
  \left(\frac{10^{-5} {\rm eV}}{m_a}\right), ~~~~~~~~ D_{\odot}\simeq 1.5\cdot 10^{13} ~{\rm cm},  
  \ee
  where we assume that the axion's energy  when the antinuggets get annihilated is slightly relativistic  $E_a\simeq 1.2 m_a$, but never becomes very relativistic, see  precise estimates below.

  The axion flux (\ref{axions})  should be compared with the flux computed in \cite{Andriamonje:2007ew}  as a result of  the Primakoff effect:
  \be
  \label{Primakoff}
  \Phi_a (\rm Primakoff)\simeq  3.75 \cdot 10^{11}\frac{g_{10}^2}{\rm cm^2 ~s}, ~~~ g_{10}\equiv g_{a\gamma}/10^{-10} GeV^{-1}, ~~~ \la E\ra=4.2~ keV.
  \ee 
The axion flux  (\ref{axions}) is much larger than the conventional flux (\ref{Primakoff}). However, the energies of the axions in these two mechanisms are drastically different. Therefore, the energy flux of the conventional flux   (\ref{Primakoff}) is also much larger than the axion energy flux due to the nuggets, 
\be
\label{energy-flux}
m_a \Phi_{\rm axions}\sim 3\cdot 10^{11}\frac{\rm eV}{\rm cm^2 ~s}, ~~~~~~~~~~   \la E\ra\Phi_a (\rm Primakoff)\sim g_{10}^2   \cdot 10^{15}\frac{\rm eV}{\rm cm^2 ~s}.
\ee
It is very instructive to compare these fluxes with conventional cold dark matter galactic axion contribution  assuming the axions  saturate the observed DM density:
\be
\label{galactic-axion}
m_a \Phi({\rm galactic ~axions})\sim  \rho_{\rm DM}\cdot v_{\rm DM}\simeq \frac{0.3~{\rm GeV}}{\rm cm^3}v_{\rm DM}\simeq 10^{16} \frac{eV}{\rm cm^2 ~s}. 
\ee

We emphasize that the estimate (\ref{axions_rate}) in this framework is almost model-independent expression as it is directly linked  to the observed EUV luminosity  (\ref{estimate}) and (\ref{total_power}). This intimate  relation between EUV luminosity and the axion luminosity emerges as a result that  both radiations 
(EUV and the axions) are related to the same physics and occur as a result of the annihilation of the antinuggets in the corona in the AQN framework. In contrast, the estimate (\ref{axions}) is a model-dependent result as it is based on our computations of the spectral properties of the axion emission
determined by the average $E_a\simeq 1.2 m_a$. The corresponding estimate   is the subject of the next subsection.

\subsection{Spectral properties}
\label{sec:4.2 Spectral properties}
As mentioned earlier in this section, study in \cite{Ge:2017idw} indicates that the total energy of an AQN finds its minimum when the axion (domain wall) contributes 1/3 of its total mass. Now consider an AQN loosing its mass when entering the solar corona, such that the axion quickly increases its portion of total mass in comparison with the equilibrium value. One should comment here that the axion domain wall in the equilibrium does not emit any axions as a result of pure kinematical constraint: the static domain wall axions are off-shell non-propagating axions. The time dependent perturbation obviously changes this equilibrium configuration.  
In other words,  the  configuration becomes    unstable with respect to emission of the axions because the total energy is no longer at its minimum. To retrieve its ground state, an AQN will therefore intend to lower its domain wall portion of the energy  by radiation of axions. To summarize: the emission of axions is an inevitable consequence during the annihilation of antinuggets in the solar corona simply for the reason to maintain the AQN stability.

\begin{figure}
    %\centering
   % \begin{subfigure}[b]{0.9\textwidth}
        \includegraphics[width=\linewidth]{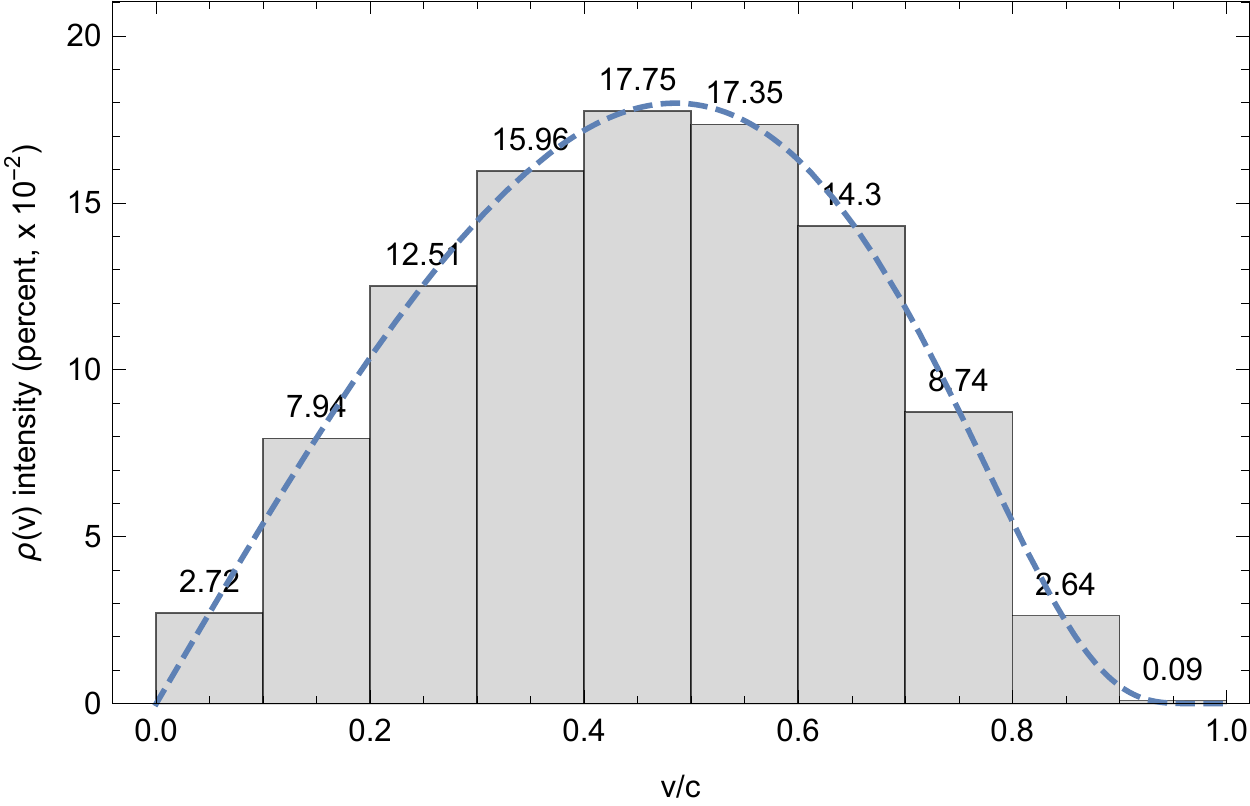}
        \caption{The normalized spectrum $\rho(v)$ vs $v/c$. The numbers inside the spectrum give the estimated percentage of the intensity per velocity BIN of 0.1c.}
        \label{fig:A percent_spectrum_velocity_full}
 %   \end{subfigure}
    %~ %add desired spacing between images, e. g. ~, \quad, \qquad, \hfill etc. 
       \end{figure}

Now, we want to identify a precise  mechanism which produces the on-shell freely propagating axions emitted by  the axion domain wall.   In this section we overview the basic idea   of the computational technique to be used. We   refer to Appendix \ref{Appendix:A Technical details. 1D approximation for the axion domain wall} for the technical details of the computations.  To address this question, we consider the general form of a domain wall solution:
\begin{equation}
\label{eq:4.2 phi soln}
\phi(R_0)=\phi_w(R_0)+\chi
\end{equation}
where $R_0$ is the radius of the AQN, $\phi_w$ is the classical solution of the domain wall, while  $\chi$ describes the excitations due to the time dependent perturbation.
We should note that, $\phi_w$ is clearly off-shell classical solution, while  $\chi$ describes the on-shell propagating axions.  Thus, whenever the domain wall is excited, namely $\chi\neq0$, freely propagating axions may be produced and emitted by the excitation modes. 

Suppose an AQN is traveling in vacuum where no annihilation event takes place, we expect the solution stays in its ground state $\phi(R_0)=\phi_w(R_0)$ which corresponds to the minimum energy state. Since there is no excitation (i.e. $\chi=0$), no free axion can be produced. However, the scenario   drastically changes when some baryon charge from the AQN get annihilated. Due to these annihilation processes, the AQN starts to loose a small amount of its mass, and consequentially its size shrinks from $R_0$ to a slightly smaller radius $R_{\rm new}=R_0-\Delta R$. Note that its quantum state $\phi(R_0)=\phi_w(R_0)$ is no longer the ground state, because a lower energy state $\phi_w(R_{\rm new})$ with lower value of $B$ becomes available. Then, we may write the current state of the domain wall as $\phi(R_0)=\phi_w(R_{\rm new})+\phi'(R_{\rm new})\Delta R$, so the domain wall now has a nonzero exciting mode $\chi=\phi'(R_{\rm new})\Delta R$ and free axions can be produced during oscillations of the domain wall. 

To reiterate: the  annihilation of antinuggets in the solar corona forces the surrounding domain wall to oscillate.  These oscillations of domain wall generate excitation modes and ultimately lead to radiation of the propagating axions.

We conclude this section by highlighting the results of the computations.  We present our results in this  section for the velocity distribution.  Equivalently, one can represent the same information in terms of the energy and momentum distribution. We refer to   Appendix \ref{Appendix:A Technical details. 1D approximation for the axion domain wall} for  the corresponding technical details.
 
 Normally, it is convenient to express the normalized spectrum as a function of the speed of emitted axion $v_a/c$, defines as follows
  \be
  \label{normalization}
 \rho(v_a)\equiv\frac{1}{\Phi^{\rm tot}_{\rm axions}}\frac{d}{dv_a}\Phi_{\rm axions}(v_a), ~~ \Phi^{\rm tot}_{\rm axions}\sim  3\cdot 10^{16}\frac{1}{\rm cm^2 ~s}
  \left(\frac{10^{-5} {\rm eV}}{m_a}\right), ~~ \int^1_0 dv_a \rho(v_a) =1.~~~~
 \ee
In Fig. \ref{fig:A percent_spectrum_velocity_full}, the we represent  the  results in the entire region of the allowed kinematical domain, from $v_a/c=0$ to 1.    We can see the distribution is roughly Gaussian, and  it  peaks near $v_a/c\sim0.5$ with  a sharp cut at $v/c\gtrsim0.9$. Thus we would expect the axion flux is relativistic but not ultra-relativistic, in contrast with conventional Primakoff effect (\ref{Primakoff}).  
   
   We should comment here that at very small velocities $v_a\ll c$ the spectrum shows the linear dependence on $v_a$.
  We think that it is  an artifact of our computational approximations. Indeed, our technical  derivation is based on the   ``thin-wall approximation'', see Appendix \ref{Appendix:A Technical details. 1D approximation for the axion domain wall}. That is, assuming the conventional thickness of the domain wall $\sim m_a^{-1}$, the thin-wall approximation is justified when $\lambda_a\lesssim m_a^{-1}$ ($\lambda_a$ stands for de Broglie wavelength of the emitted axion). While this condition is marginally satisfied in case of an relativistic axion, it is badly violated in non-relativistic regime.  A different computational technique is obviously required to address the question about the spectral properties in non-relativistic regime.
  The region of small $v_a\leq 0.1c$  contributes very little (around $\sim 1\%$) to the normalization integral (\ref{normalization}).  Therefore, it will  be ignored for the purposes of the present  work.

 \section{Design: CAST and other dark matter axion antennae   \label{design}}
 In this section we discuss the discovery potential   of the  relativistic axions   emitted  by the  AQNs as a result of the annihilation events. We consider separately the solar axions in subsection \ref{sun} and the axions radiated in the Earth's atmosphere and in the Earth's core in subsection \ref{earth}. 
   \subsection{\label{sun}AQNs in   the Sun}
The reasoning about this new production of the DM axions of solar origin emerging during the interaction of AQNs with the outer Sun along with the derived velocity spectrum, define their detection scheme. We take here as an example the CAST-CAPP DM axion antenna \cite{CAPP:2018} whose commissioning is scheduled soon. In fact, this antenna is of the Sikivie type, whose  design has been supplemented with a fast scanning mode, becoming thus a quasi wide-band DM axion antenna. This modification makes CAST sensitive to streaming DM axions including axion mini-clusters(aMCs), without compromising the conventional search for galactic axions \cite{ZIOUTAS:2017str, FSZ:2017}.
 
The DM axions from AQN defragmentation in the corona are the same axions with the same coupling constant $f_a$ 
which is the subject of many other axion  searches, see reviews \cite{vanBibber:2006rb, Asztalos:2006kz,Sikivie:2008,Raffelt:2006cw,Sikivie:2009fv,Rosenberg:2015kxa,Marsh:2015xka,Graham:2015ouw,Ringwald:2016yge}.
However, there are  few differences we want to   point out here because these distinct features  are important for their detection and identification:
 
1) the axions which are produced due to the AQN annihilations in the Sun     will be 
emitted from narrow transition region with width of few hundred kilometres at the altitude around 2000 km where the most of annihilation events occur, see Fig. 9 in  ref.\cite{Raza:2018gpb}.
The probability for these  axions to be reabsorbed inside the Sun is negligible, similar to the conventional arguments  \cite{Andriamonje:2007ew} for the axions produced by  the Primakoff effect.     The sensitivity  of  available    instruments is not likely to resolve this   structure at present time. Therefore,   as a first rough approximation for the Earth's observable one can ignore this structure and assume that   entire Sun  with angular diameter of $0.55^o$  emits  these axions.

2) these axions are characterized by the  relativistic velocities ($v \approx 0.5$ c), see Figure \ref{fig:A percent_spectrum_velocity_full}. The corresponding distribution   is very distinct  from the galactic axions with typical velocities of order $v \approx 10^{-3}$ c. 
 
3) if the AQNs are at the origin of spatiotemporally confined solar activity \cite{Zhitnitsky:2017rop}, this will provide a trigger in real time due to the continuous monitoring of the Sun by various observatories. This is unique in the field of the dark sector, and, with  Figure \ref{fig:A percent_spectrum_velocity_full} in mind, the warning time will be at least 15 minutes. In addition, this implies an improvement of the signal-to-noise ratio, allowing also to distinguish the solar DM component from the galactic one ($v \approx 10^{-3}$ c).
 
In what follows we would like to consider a specific example of the planning experiments \cite{CAPP:2018} which in principle are capable 
to detect  these axions produced by AQN mechanism. In our estimates below we use the  technical characteristics as presented in  \cite{CAPP:2018}.
The basal solar DM axion flux at the Earth is about 1 per mille   compared to that of the galactic axions. 
The size of the CAST-CAPP cavities require that the de Broglie wavelength of the axions be $\lambda_{dB} \sim 4 $ meters, ensuring phase matching between several cavities. This condition applies to many axion DM experiments  (e.g., \cite{CAPP:2018}). The various CAPP cavities can cover an axion rest mass range of about 3 to 120 $\mu $eV (0.7 to 30 GHz). The CAST search will start at about 20 $\mu $eV, being thus sensitive to velocities below 0.01c. 

One should mention that the DM stream can drastically change the number of   annihilating events in the solar atmosphere as argued in \cite{Zioutas}. The corresponding changes will lead to drastic temporary variation of the axion flux from the Sun. These drastic changes can be anticipated as the time delay is at least 15 minutes.
It gives us a hope to observe such time-dependent short enhancements in the axion flux.

At the same time a planetary gravitationally focused DM stream of AQNs can change the number of   annihilating events in the solar or earth atmosphere by a factor up to $10^6$, see refs.\cite{Patla:2014,Zioutas:2016,ZIOUTAS:2017str}.  Interestingly, in recent work \cite{Turushev:2017} an amplification factor by as much as $1.2\cdot 10^{11}$  is given for the Sun as gravitational lens for incident light downstream at about 520 AU. Given the fact that the Einstein Ring and the deflection angle increase with velocity $v$ as $v^{-1}$ and $v^{-2}$, respectively, all this improves the situation for an Earth observer searching for non-relativistic particles with $v\ll c$. Thus, the impact on streaming AQNs  during alignment of Earth - Sun with the assumed AQN stream is much larger than the aforementioned planetary impact towards the Sun. This is actually reasonable, i.e., the Sun is of course a better gravitational lens than any planet. Therefore, an axion haloscope like CAST may well profit from such a drastic flux enhancement by the Sun. Moreover, signatures from the active Earth atmosphere could be used as axion  trigger, which in ref. \cite{Zioutas}
have shown even a planetary dependence. More intriguing might be the decades long puzzle of ionization excess in December. We recall, that on the 18-th of December there will be an alignment Earth-Sun-Galactic Center  within $5.5^o$.

The recent theoretical and numerical studies  on propagation of the nuggets  in the solar atmosphere 
   have  produced  very encouraging results \cite{Raza:2018gpb}.
 In that  work the detail  numerical simulations have been carried out.  It has been confirmed that the total energy injected into the corona as a result of the annihilation events of the AQNs with the  solar material is order of $10^{27}~ {\rm erg/s}$. This is very robust prediction of the model  in full agreement with observations (\ref{estimate}), see Fig 10 in  \cite{Raza:2018gpb}. This should be considered  as a highly nontrivial consistency check of the AQN framework because  the original Monte Carlo  sample was around  $\sim 10^{10}$ particles distributed up to distances $\sim10~{\rm AU}$. Furthermore, the  annihilation processes effectively started  at the altitude $\sim 2200~ {\rm km}$ which precisely corresponds to the transition region where  some  drastic changes  are known to occur. The most of the energy is deposited in the transition region, see  Fig 9 in  \cite{Raza:2018gpb}, which implies that the axions will be released at the same time at the altitude around $\sim 2200~ {\rm km}$.

As we mentioned at the end of section \ref{sec:4.2 Spectral properties} the present  computations at small velocities $v_a\ll c$ 
(which are required to analyze the aforementioned enhancements related to the gravitationally focused DM stream)
are not reliable. Therefore, we leave the corresponding estimates with  possible enhancement factors  for future studies. 
The only comment we would like to make to conclude this subsection   is as follows. As we argued above the streaming dark matter   axions may be the better source for their discovery than the widely assumed isotropic DM. This is because, a large axion flux enhancement can take place, temporally, due to gravitational lensing when the Sun and/or a planet are aligned with the stream or an axion caustic pointing to the Earth.

 \subsection{\label{earth}AQNs in  the  Earth}
 In this subsection we want to make few comments on differences between the AQNs propagation in   the solar atmosphere in comparison with the    Earth's atmosphere. The drastic changes between the two systems have been previously discussed in \cite{Zhitnitsky:2017rop}. 
 From the theoretical viewpoint the solar atmosphere is much simpler system which is  easier  to study than the Earth's atmosphere.  The basic reason for such simplification is that the solar corona is a highly ionized system consisting  mostly  protons and electrons. It should be contrasted with Earth's atmosphere  where some atoms (mostly heavy elements $N$ and $O$)  are neutral and some are partly ionized. 
 The interaction of these heavy neutral elements  with the AQNs is a highly complicated problem as  the most likely outcome of the collision is the elastic reflection  rather than penetration deep inside the nugget with some probability of partial annihilation processes, which inject the energy and produce the axions.  The corresponding enhancement factor in the Sun due to long range Coulomb  forces in highly ionized plasma at temperature $T\simeq 10^6$ K was parametrized 
 in \cite{Raza:2018gpb} by effective size  $R_{\rm eff}\gg R$ to be distinguished from    its physical size $R$. This implies that the effective cross section for protons with AQN in the Sun is approximated as $\sim \pi R_{\rm eff}^2$ while 
 a similar cross section for neutral atoms is $\sim \pi R^2$.

As we mentioned above, similar computations have not been carried out  for Earth's atmosphere yet.
 Nevertheless,  we would like to make here few estimates for the axion flux due to the disintegration of the AQNs in the  Earth's atmosphere. 
 The AQN flux on Earth is estimated as follows
\begin{equation}
\label{eq:flux}
\frac{dN}{dA ~ dt} = n_{\rm AQN}v  \approx 0.3\cdot \left( \frac{10^{25}}{\la B\ra} \right) {\rm km}^{-2} {\rm yr}^{-1}, ~~~~ n_{\rm AQN}\simeq \frac{\rho_{\rm DM}}{m_p \la B\ra}.
\end{equation}
  This tiny rate  represents the main reason why the direct detection  of the nuggets requires  detectors with large area such as 
  Pierre Auger Observatory or  Telescope Array to observe the showers produced by the AQNs entering the atmosphere \cite{Lawson:2010uz,Lawson:2012vk}.

The annihilation processes are much less efficient in Earth's atmosphere than in the Sun as mentioned above due to the drastic difference between  $R$ and $R_{\rm eff}$ in the ionized hot plasma. It has been estimated in \cite{Lawson:2010uz,Lawson:2012vk,Lawson:2013bya} that only small portion of the AQN's   mass $\Delta M\simeq 10^{-10} {\rm kg}$ will get annihilated in the Earth's atmosphere. This represents only tiny portion $\sim \Delta B/B \sim 10^{-8}$ of a typical nugget which can get annihilated in the atmosphere.
On entering the earth's crust the nugget will continue to deposit energy along 
its path, however this energy is dissipated in the  surrounding rock and is unlikely 
to be directly observable. Generally the nuggets carry sufficient momentum to travel 
directly through the earth and emerge from the opposite side.  However a finite  fraction of the AQNs  ($\Delta B/{B}$ could be around 10$\%$)
may be captured by the Earth and deposit all their energy in dense regions\footnote{A better estimate requires a precise numerical simulations as different nuggets have different baryon charges $B$, different sizes, different impact velocities.  Furthermore, some of the nuggets will have very short path, much shorter than the Earth's radius $R_{\oplus}$, while other nuggets will have   long  paths of order $\sim 2R_{\oplus}$, depending on the angle of the impact. }.

While the observation  of the $E\&M$ showers due to the nuggets entering the Earth's atmosphere indeed requires very large area detectors, the observation of the axions (which have been produced as a result of the annihilation events in the very deep underground) is possible, and in fact very promising. Indeed, 
the corresponding axion flux  can be estimated from (\ref{eq:flux}) as follows
\be
\label{earth-axion}
m_a \Phi ({\rm Earth ~axions})\sim \frac{(2 \Delta B)~{\rm GeV}}{3}\cdot \frac{dN}{dA ~ dt}\sim  10^{16}\cdot \left(\frac{\Delta B}{B}\right)\frac{\rm eV}{\rm cm^2 ~s}, 
\ee
   where we assume that $\Delta B\sim B$ charges  of the AQNs  get annihilated in the earth's core, and each event of annihilation produces $2~ \rm GeV$ energy deep underground.
   The key point here is that $\sim 1/3$ of this energy is radiated in the form of axions similar to our arguments leading to (\ref{axions_rate}).
   Furthermore, these axions will have the typical velocities $v\sim 0.5 c$ as plotted on Fig. \ref{fig:A percent_spectrum_velocity_full}.
   These axions, in principle, can be observed, in contrast  with conventional  $E\&M$ energy which is dissipated in the  
   mantle or in the earth's core and completely lost for the direct observations.  Notably, for the calculation of (\ref{earth-axion}) we have not taken into account the aforementioned temporally flux enhancement due to planetary (up to $\sim 10^6$) and in particular solar gravitational focusing (up to 
$\sim 10^{11}$). 
   
   Interestingly, the axion flux  (\ref{earth-axion})   which is generated due to the
   AQN annihilation events is much larger than the  flux (\ref{energy-flux}) generated due to the AQN annihilation events in the solar corona. 
   At the same time, the axion flux  (\ref{earth-axion}) is the same order of magnitude 
   as  the conventional cold dark matter galactic axion contribution (\ref{galactic-axion}). This is because  the parameter 
  ${\Delta B}/{B}\sim 1$ is expected to be order of one, as  a finite  portion of the AQNs  will get annihilated in the Earth's core.  However, the 
   wave lengths of  the axions produced due to AQN annihilations, are much shorter  due to their  relativistic velocities  $v\sim 0.5 c$, in contrast with conventional galactic isotropic axions with $v\sim 10^{-3}c$. Therefore, these two distinct contributions can be easily discriminated.

\section{Conclusion and future directions\label{conclusion}}
    The main claim of this work is that there is a  new mechanism  of the axion production (Figure 1)  with drastically different spectral features (in comparison with conventional galactic axions characterized by  $v\sim 10^{-3}c$ and the solar axions characterized by  typical energies $\la E\ra\simeq 4~ $keV). The corresponding fluxes are given by eqs. (\ref{earth-axion}) for the earth's core and  (\ref{energy-flux}) for the solar corona correspondingly.  The new mechanism is entirely based on the non-orthodox AQN dark matter  model. 
    
    Why this new AQN framework (and  accompanying  the axion emission)  should be taken seriously? We refer to Section 
    \ref{sec:QNDM} for overview of this DM model. The only comment we would like to make here is that this model was invented long ago as a natural 
    explanation of the observed ratio  (\ref{Omega}) between visible and dark matter 
densities. This model passed all the tests to be qualified as a valid DM candidate. In context of the present work the most important 
    feature of this model is that it   may potentially resolve the old renowned puzzle (since 1939) known in the community under the name ``the Solar Corona Mystery".  In particular, this model, without adjusting any parameters, generates the observed EUV luminosity (\ref{estimate})
    as recent numerical Monte Carlo simulations show \cite{Raza:2018gpb}. The corresponding intensity $\sim 10^{27}~ {\rm erg/s}$ is 
    entirely determined by the dark matter density in the solar system. The mysterious temperature inversion around 2000 km also finds its natural explanation as the most of AQNs inject their energy in this transition region, see Fig.9 in \cite{Raza:2018gpb}. Surprisingly, the new axion production leading to the estimate (\ref{energy-flux}) occurs precisely at  this transition region.   
    
 Following the numerical estimates  given above, most DM axion searches (which are presently running, or planning to start taking the data in near future)
 have the potential to discover DM axions from AQNs
which are produced in the solar corona (\ref{energy-flux}) or in the earth's core (\ref{earth-axion}). The key element which discriminates these axions from conventional DM galactic axions is their wide relativistic velocities  at $v\sim 0.5 c$.
Therefore, in the present work we choose not to specify the instruments which are most suitable and capable for such an analysis. Generically speaking  the proper  instruments  must be either wide-band by default, or, they must implement the fast scanning mode \cite{ZIOUTAS:2017str, FSZ:2017, Zioutas:2016}
.
 
In addition, following the planetary dependence of the atmospheric electron density \cite{Zioutas}, a similar dependence also for a putative DM axion signature from defragmented AQNs seems possible. Such a dependence can be used for signal identification as well as for improving the signal-to-noise ratio resulting to a better detection sensitivity.

\section*{Acknowledgements} 
The work of A. Z. and X. L. was supported in part by the Natural Sciences and Engineering Research Council of Canada. H.F., Y.S. and K.Z. would like to thank the CAST Collaboration for inspiring this work. K. Z. thanks Slava Turyshev for encouraging exchanges. K. Z. is partly funded by GSRT Athens/Greece \&ELKE/University of Patras/Greece. The work of Y. S. was supported by IBS- R017-D1 of the Republic of Korea.

\appendix
\section{Technical details. Axion emission from the domain wall}
\label{Appendix:A Technical details. 1D approximation for the axion domain wall}
 In this Appendix we want to study the spectral properties of the axion's emission as a result of time-dependent perturbations of the axion domain wall.
 We want to focus on the axion portion of the axion DW, which also includes other fields such as $\pi, \eta'$,  see \cite{Forbes:2000et}. It   also contains a phase describing the baryon charge distribution on the surface of the nugget as discussed in 
   \cite{Liang:2016tqc}. Exact features of the profile functions for all these fields  are not important for our purposes. Therefore, one can simplify our computations by considering the  following  
    effective Lagrangian with two  degenerate vacuum states\footnote{In our previous studies  \cite{Liang:2016tqc,Ge:2017ttc,Ge:2017idw} we always discussed the so-called $N=1$ domain walls. It implies that   the vacuum is unique and the DW solution interpolates between one and the same physical vacuum. This interpolation always occurs as a result of variation of the axion field together  with another fields, such as $\pi$ or $ \eta'$ as discussed in \cite{Forbes:2000et}. These additional fields do not generate much changes in  the domain wall tension, nor they 
    affect our analysis of the axion production, which is the subject of the present work. Therefore, we ignore these fields to simplify notations and qualitative analysis in this work.}.  
\begin{equation}
\label{eq:A L(phi)}
{\cal S}[\phi]
=\int d^4x\left[\frac{1}{2}(\partial_\mu\phi)^2
-\frac{g^2}{4}\left(\phi^2-\frac{\pi^2}{4}f_a^2\right)^2\right],
\end{equation}
where $g=\frac{\sqrt{2}}{\pi}\frac{m_a}{f_a}$, and we set the effective axion angle as $\phi/f_a\equiv\theta+{\rm arg~det}M+\pi/2$ (note that we shift the angle by $\pi/2$ for convenience of calculation). In this toy model, the domain wall solution has an exact form
\begin{equation}
\label{eq:A phi_DW}
\phi_{w}=\frac{\pi}{2}f_a \tanh\left[\frac{1}{2}m_a(z-z_0)\right]
\end{equation}
for arbitrary $z_0$. Despite the simplicity, this model contains all the essential feature of sine-Gordon Lagrangian we need in this computation. For example, the surface tension of the domain wall in this model is
\begin{equation}
\label{eq:A sigma def}
\sigma\equiv\int_{-\infty}^{\infty}(\phi_{w}'(z))^2
=\frac{\pi^2}{6}f_a^2m_a
\end{equation}
which is very similar to the classic value $\sigma\simeq8f_a^2m_a$ computed for the Sine-Gordon potential. 
\exclude{However, as mentioned in \cite{Ge:2017idw}, the ``effective'' surface tension of a nugget is actually a much smaller due to the appreciable domain wall curvature in the thin-wall approximation. Such correction may be qualitatively described as $\sigma_{\rm eff}\equiv\kappa\cdot\sigma$, where $\kappa\simeq10^{-4}$. To account for this effect in our double-well model (\ref{eq:A L(phi)}), and continue to use thin-wall approximation we should replace $f_a^2\mapsto f_{a,\rm eff}^2\equiv\kappa\cdot\frac{48}{\pi^2} f_a^2$  in the final numerical estimates.
}

Our goal now is to compute the   excitations $\chi(t,z)$ in the time dependent background. These excitations will be eventually identified with  the axions emitted by the axion DW.  To achieve this task   we expand $\phi(t,z)=\phi_{w}(z-R_0)+\chi(t,z)$, which gives
\begin{equation}
\label{eq:A S2}
{\cal S}[\phi]
={\cal S}[\phi_{w}]
+\int dt\int d^2x_\perp\int dz \left[\frac{1}{2}\dot{\chi}^2
-\frac{1}{2}\chi L_2\chi\right]+{\cal O}(\chi^3),
\end{equation}
where $L_2$ is a linear differential operator of the second order,
\begin{equation}
\label{eq:A S2_ass L2}
\begin{aligned}
L_2
&=\left.\left[-\frac{\partial^2}{\partial z^2}
+2g^2\phi^2+g^2(\phi^2-v^2)\right]\right|_{\phi=\phi_{w}(z-R_0)} \\
&=-\frac{\partial^2}{\partial z^2}
+\frac{m^2}{2}\left[3\tanh^2\left(\frac{1}{2}m_a(z-R_0)\right)-1\right].
\end{aligned}
\end{equation}
The corresponding equation of motion is therefore
\begin{equation}
\label{eq:A PDE}
\frac{\partial^2}{\partial t^2}\chi=-L_2\chi.
\end{equation}
To look for the initial conditions, we now want to describe the emission of axions in one cycle of oscillation. As mentioned in Sec. \ref{sec:4.2 Spectral properties}, annihilation of baryon charge results in oscillations of domain wall. Assuming the oscillation is approximately adiabatic, it is sufficient to only analyze the first half of an oscillation -- say, the ``contraction period''-- where the domain wall shrinks from $R_0$ to a slightly smaller size $R_0-\Delta R$. We assumed the rest half of the cycle, the ``expansion period'', is just the time-reversed and produces an equivalent contribution. We may write down such initial conditions as
\begin{subequations}
\label{eq:A phi_IC}
\begin{equation}
\label{eq:A phi_IC1}
\begin{aligned}
\phi(0,z)=\phi_w(z-R_0)
\end{aligned}
\end{equation}
\begin{equation}
\label{eq:A phi_IC2}
\begin{aligned}
\quad\phi(\frac{1}{2}t_{\rm osc},z)=\phi_w(z-R_0+\Delta R)+({\rm excitations})
\end{aligned}
\end{equation}
\end{subequations}
where $t_{\rm osc}$ denotes the period of one full oscillation. The excitation modes in condition \eqref{eq:A phi_IC2} is unknown and depends on the conversion rate from excitation modes to freely propagating axions. In terms of $\chi$, the initial conditions \eqref{eq:A phi_IC} imply
\begin{subequations}
\label{eq:A chi_IC}
\begin{equation}
\label{eq:A chi_IC1}
\begin{aligned}
\chi(0,z)=0
\end{aligned}
\end{equation}
\begin{equation}
\label{eq:A chi_IC2}
\begin{aligned}
\chi(\frac{1}{2}t_{\rm osc},z)
\simeq\sqrt{\eta}\phi_w'(z-R_0)\Delta R+{\cal O}(\Delta R^2),
\end{aligned}
\end{equation}
\end{subequations}
where we introduce a free parameter $\eta$ to account for the conversion rate to axion radiation, so $\eta$ varies from 0 to 1. An efficient conversion corresponds to $\eta\sim1$, and a poor rate of conversion corresponds to $\eta\ll1$. In general, we should expect $\eta\sim1$.
This numerical factor does not modify our conclusion about the spectrum.  It may only affect the intensity which is   fixed by the observed EUV emission,  and it is given by eq.  (\ref{axions}).

We now express $\chi$ in term of normalized basis
\begin{equation}
\label{eq:A chi in a chi_p}
\chi(t,z)
=\int_{-\infty}^\infty dp~a_p(t)\chi_p(z),\quad
\chi_p(z)\equiv\frac{1}{\sqrt{4\pi E_a S}}e^{ipz}.
\end{equation}

Note that $L_2$ is diagonal in the basis $\chi_p$
\begin{equation}
\label{eq:A L_2 in chi_p}
\begin{aligned}
\int d^3x\chi_p^*(z)L_2\chi_q(z)
&=\frac{1}{4\pi E_a}\int_{-\infty}^{\infty}dz~
e^{-i(p-q)z}\left\{
q^2+\frac{m_a^2}{2}\left[3\tanh^2\left(\frac{1}{2}m_a(z-R_0)\right)-1\right]
\right\}  \\
&=\frac{p^2}{2E_a}\delta(p-q)
+K_{p,q}\frac{m_a^2}{4\pi E_a} e^{-i(p-q)R_0}
\int_{-\infty}^\infty dz~e^{-i(p-q)z} \\
&=\frac{\delta(p-q)}{2E_a}(p^2+K_{p,q}m_a^2)
\end{aligned}
\end{equation}
where in the intermediate step, we have defined the ratio
\begin{equation}
\label{eq:A L_2 in chi_p_ass Kpq}
K_{p,q}
\equiv\frac
{\int_{-\infty}^{\infty}dz~e^{-i[p-q-{\rm sign}(z)\cdot i\varepsilon]z}
\frac{1}{2}\left[3\tanh^2\left(\frac{1}{2}m_a z\right)-1\right]}
{\int_{-\infty}^{\infty}dz~
e^{-i[p-q-{\rm sign}(z)\cdot i\varepsilon]z}}
\end{equation}
for simplicity of calculation. Note that $K_{p,q}$ is finite and well defined in the entire range of $p,q$. For our computations when $p=q$ the parameter $K_{p,q}=1$ as  $K_{p,q}\delta(p-q)=\delta(p-q)$.  Then Eq. \eqref{eq:A L_2 in chi_p} is simplified to
\begin{equation}
\label{eq:A L_2 in chi_p_2}
\begin{aligned}
\int d^3x\chi_p^*(z)L_2\chi_q(z)
=\frac{\delta(p-q)}{2E_a}(p^2+m_a^2).
\end{aligned}
\end{equation}
Our original equation  \eqref{eq:A PDE} now can be simplified into
\begin{equation} 
\label{eq:A eqns for a_p final}
\frac{d^2}{dt^2}a_p=-E_a^2(p)a_p,\quad
E_a(p)\equiv \sqrt{p^2+m_a^2}.
\end{equation}

Solving Eq. \eqref{eq:A eqns for a_p final} with initial conditions \eqref{eq:A chi_IC} give
\begin{equation}
\label{eq:A a_p soln Final}
a_p(t)
=e^{-ip R_0}\Delta R\frac{\pi}{2}\frac{f_a}{m_a}\sqrt{4\pi\eta SE_a}
\frac{\sin E_at}{\sin(\frac{1}{2} E_at_{\rm osc})}
p~{\rm csch}\left(\frac{\pi p}{m_a}\right).
\end{equation}
Then, the total radiation energy of the domain wall is obviously
\begin{equation}
\label{eq:A energy density}
\begin{aligned}
E_{\rm rad}
&=\int d^3x~\frac{1}{2}\chi^*
\left[-\frac{\partial^2}{\partial t^2}+L_2\right]\chi  
=\int_{-\infty}^\infty dp~\frac{1}{2}E_a(p) |a_p|^2  \\
&=\int_{m_a}^\infty dE_a~\pi^3\eta S \Delta R^2\left(\frac{f_a}{m_a}\right)^2
\left[\frac{\sin E_a t}{\sin(\frac{1}{2}E_a t_{\rm osc})}\right]^2
E_a^3p~{\rm csch}^2\left(\frac{\pi p}{m_a}\right).  \\
\end{aligned}
\end{equation}
More generally, the domain wall is oscillating in a shallow cavity $S\Delta R$, so the excitation energy density (in volume) is $E_{\rm rad}/S\Delta R$. Then the axion flux spectrum $\Phi_{\rm rad}$ emitted from a single AQN is clearly
\begin{equation}
\label{eq:A flux spectrum}
\begin{aligned}
\frac{1}{S}\frac{d}{dE_a}\Phi_{\rm rad}
=\frac{p}{E_a^2}\frac{d}{dE_a}\left(\frac{E_{\rm rad}}{S\Delta R}\right)
=\pi^3\eta \Delta R\left(\frac{f_a}{m_a}\right)^2
\left[\frac{\sin E_a t}{\sin(\frac{1}{2}E_a t_{\rm osc})}\right]^2
E_a p^2{\rm csch}^2\left(\frac{\pi p}{m_a}\right).
\end{aligned}
\end{equation}
Such spectrum indicates an average energy $\langle E_a\rangle=1.18m_a$.  One may also see Figs. \ref{fig:A percent_spectrum_energy and momentum}, where the normalized flux spectra as a function of $E_a$ and $p$ are plotted in Figs. \ref{fig:A percent_spectrum_energy and momentum}. 
\begin{figure}
    \centering
    \begin{subfigure}[b]{0.45\textwidth}
        \includegraphics[width=\textwidth]{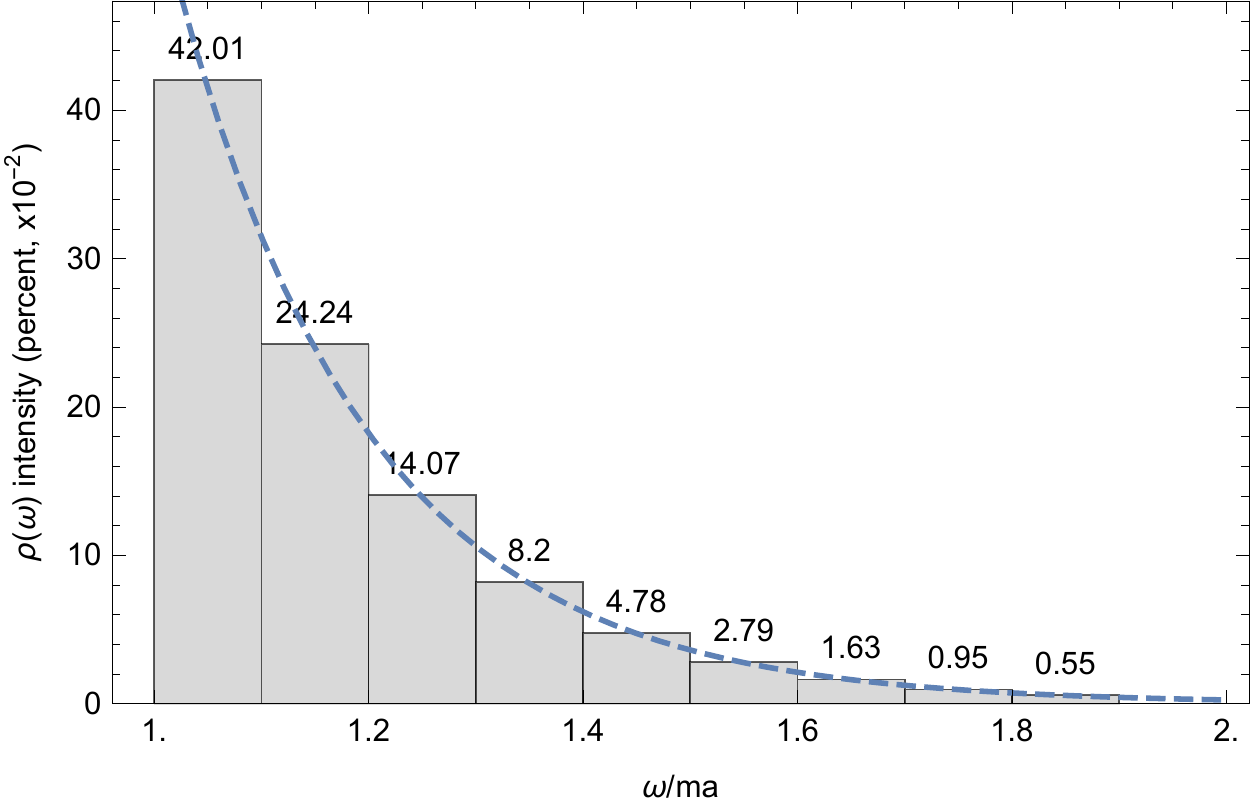}
        \caption{The normalized spectrum $\rho(E_a)$ vs $\omega$.}
        \label{fig:A percent_spectrum_energy}
    \end{subfigure}
    %~ %add desired spacing between images, e. g. ~, \quad, \qquad, \hfill etc. 
    \begin{subfigure}[b]{0.45\textwidth}
        \includegraphics[width=\textwidth]{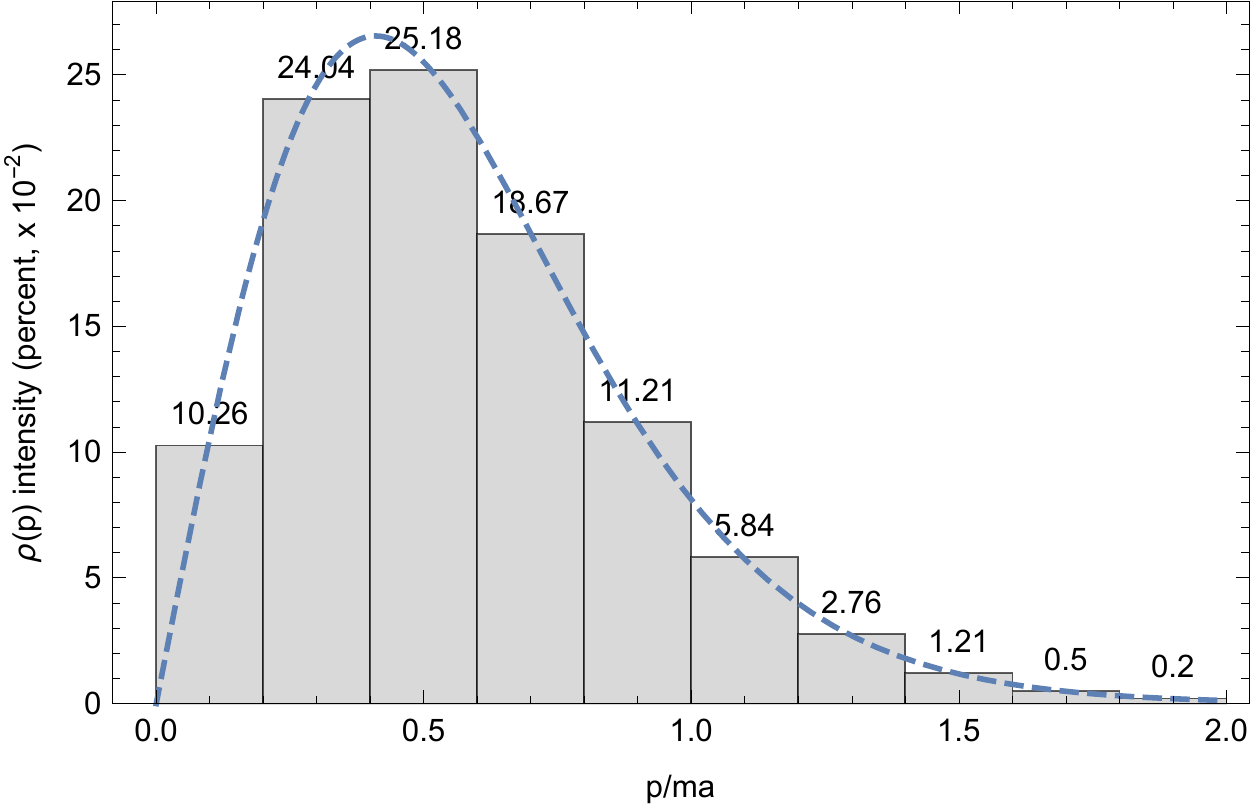}
        \caption{The normalized spectrum  $\rho(p)$ vs $p$.}
        \label{fig:A percent_spectrum_momentum}
    \end{subfigure}
    \caption{(Normalized) flux spectrum for energy and momentum.}
    \label{fig:A percent_spectrum_energy and momentum}
\end{figure}
It is also useful to obtain the spectrum as a function of axion velocity $v_a$, 
\begin{equation}
\label{eq:A flux spectrum_velocity}
\begin{aligned}
\frac{1}{S}\frac{d}{dv_a}\Phi_{\rm rad}
&=\pi^3\eta \Delta R\left(\frac{f_a}{m_a^2}\right)^2
\left[\frac{\sin E_a t}{\sin(\frac{1}{2}E_a t_{\rm osc})}\right]^2
E_a^3 p^3{\rm csch}^2\left(\frac{\pi p}{m_a}\right),
\end{aligned}
\end{equation}
which gives Figs. \ref{fig:A percent_spectrum_velocity_full} in Sec. \ref{sec:4.2 Spectral properties}.

We conclude this appendix with the following comments. The main goal  of this analysis  is  the computation of the spectrum which is plotted above. 
The intensity of the radiation is determined  by our expressions (\ref{axions_rate}) and (\ref{axions}) which are based on assumption that 
the travel time of the AQN in the solar corona is sufficiently   long. Therefore,  the total charge of the antinuggets  will be completely annihilated, and   the total intensity of the axion emission is fixed and given by (\ref{axions_rate}) and (\ref{axions}). 
The numerical analysis carried out in \cite{Raza:2018gpb} supports this assumption as most of the nuggets indeed get annihilated 
at the altitude around 2000 km. 

The analysis presented above suggests that the typical  velocities of the emitted axions $v_a\simeq 0.5c$. This is an expected result   because the energies of the emitted axions are determined by the moving domain wall which normally have velocities close to the speed of light, i.e. $\Delta R\sim t_{\rm osc}$. Precisely this condition eventually determines the spectral density of the emitted axions. 

The basic picture of the emission developed in this Appendix is based on thin wall approximation when infinitely  large (along $x,y$ directions) DW 
moves with acceleration and    emits axion waves moving along $z$ direction. It is quite obvious that this approach is not justified when 
 de Broglie wavelength $\lambda_a$ is comparable   with the thickness of the domain wall as explained in Section \ref{sec:4.2 Spectral properties}.
 It implies that the small velocity portion of the spectrum may receive large corrections as a result of break down of the thin wall approximation.
 Linear dependence on the  velocity at small  $v_a\ll c$ is manifestation of this approximation when the system can be shifted along $x,y$ directions
 without changing the system.

%\newpage

\end{document}